\documentclass[journal=jpclcd,manuscript=letter]{achemso}
\usepackage{graphicx}
\usepackage{amsmath}
\usepackage{amssymb}
\usepackage{xcolor}
\usepackage{bm}
\usepackage{enumitem}

\author{Jos\'e L. Movilla} 
\affiliation{Dept. d'Educaci\'o i Did\`actiques Espec\'ifiques, Universitat Jaume I, 12080, Castell\'o, Spain}
\email{movilla@uji.es}
\author{Josep Planelles}
\affiliation{Departament de Qu\'{\i}mica F\'{\i}sica i Anal\'{\i}tica,
Universitat Jaume I, E-12080, Castell\'o de la Plana, Spain}
\author{Juan I. Climente}
\affiliation{Departament de Qu\'{\i}mica F\'{\i}sica i Anal\'{\i}tica,
Universitat Jaume I, E-12080, Castell\'o de la Plana, Spain}

\date{\today}

\title{Excitonic structure in CsPbBr$_3$ nanocubes, nanorods and nanoplatelets: the effect of dimensionality}


\begin{document}

\begin{abstract}
We present a theoretical study comparing the excitonic ground state properties of CsPbBr$_3$ nanocrystals
 with different dimensionality: nanorods (quasi-1D), nanoplatelets (quasi-2D) and nanocubes (quasi-3D).
 All three systems are described on an equal footing, by means of a general variational effective mass model, 
which captures the influence of quantum confinement, dielectric confinement, electron-hole correlations, 
and polaronic effects (within a Haken model).
The strongly confined directions squeeze the exciton (X) wavefunction and enhance Coulomb attractions
along the weakly confined directions.
This stimulates superradiance, causing radiative recombination rates to speed up from cubes to platelets
and to rods, in line with recent experiments.
The anisotropic local field factor is a secondary, yet non-negligible, mechanism further enhancing radiative rates.
X binding energies are also determined primarily by the directions of strong confinement, 
which is also consistent with experiments. 
Weakly confined directions, however, become influential for small aspect ratios.
Dielectric confinement plays a major role in determining the binding energies, and less so in the interparticle distances. 
For all dimensionalities, the biexciton (XX) geometry is that of a distorted tetrahedron, 
rather than squared or linear distributions that would result in Coulomb-governed 2D and 1D structures.

\end{abstract}

\begin{tocentry}
\begin{center}
\includegraphics[width=2in]{fig_toc.pdf} 
\end{center}
\end{tocentry}

\maketitle


\newpage

Metal halide perovskite nanocrystals made of CsPbBr$_3$ are materials of current interest for optoelectronics.
They combine the advantages of 3D lead halide perovskites (ease of synthesis, defect-tolerant optics, long carrier diffusion lengths)
with benefits imparted by the low dimensionality and the use of surface ligands (greater phase stability, tunability of the electronic structure).\cite{DeyACS,ShamsiCR}  
By now, progress in size and shape control has enabled the synthesis of CsPbBr$_3$ nanocrystals with all possible dimensionalities:
 quasi-1D nanorods, quasi-2D nanoplatelets and quasi-3D nanocubes, whose distinct photophysics make them potential candidates 
for different optoelectronic applications.\cite{MishraJPCL,WangOM,YingPC,LiACSami,LiNML,AftabNE}
\\

Understanding the excitonic structure of these systems is a requirement to guide further developments.
Several theoretical studies have analyzed the X fine structure,\cite{GhribiPRM,RamadeNS,TliliNL,FuNL}
binding energy and radiative recombination\cite{BlundellPRB,MovillaPRB,MovillaNSa,ZhuNAT} 
in \emph{specific} types of cuboidal CsPbBr$_3$ nanocrystals, usually either quantum dots or nanoplatelets.
These magnitudes are of paramount importance to define the optical response of any semiconductor nanocrystal.\cite{ShamsiCR}
The goal of the present work is to provide an encompassing vision, by systematically comparing
the X binding energy ($\Delta_X$) and recombination rate ($k_X^{rad}$) of nanocubes (NCs), nanoplatelets (NPLs) and nanorods (NRs).
In other words, 
 we study the influence of dimensionality on the excitonic structure of CsPbBr$_3$ nanocrystals. 
In doing so, we shall provide answers for several questions that stand in the literature. Namely:
\begin{enumerate}[label=(\roman*)]
	\item Recent fits to absorption data suggest that X and XX binding energies in NCs and NRs are set solely by the length 
		of the edge with strongest confinement.\cite{OrielNL}
		Under which conditions is this true? How influential can the weakly confined directions be? 
	\item To what extent can $\Delta_X$ be modulated through quantum and dielectric confinements in each geometry?
		Both factors scale inversely with the nanocrystal size, but they can be engineered independently,\cite{RodinaJETP,ChakraPCCP}
		which makes the question relevant for the design of practical applications.
	\item How does the coexistence of strongly and weakly confined directions influence X properties through electron-hole correlations?
	\item Recent experiments show that NRs display the fastest X radiative decay among CsPbBr$_3$ nanocrystals.\cite{ZhuNL} 
		Is this because 1D systems benefit from superradiance\cite{ZhuNAT} more than 2D and 3D ones?
		Or is it because of the anisotropic polarizability\cite{vonToperczerNR}? Can we expect NPLs or NCs to outperform NRs?
	\item How do X and XX inter-particle distances and geometry evolve with dimensionality? 
		Electrostatically, XX in 2D structures should have planar square geometry to maximize attractions and minimize repulsions.\cite{SinghPRB}
		However, computational studies on NPLs suggest that the competition between Coulomb interactions and quantum confinement leads to
		a distorted tetrahedron geometry instead.\cite{ClimenteJPCL,MaciasNS} 
		What is the situation in quasi-1D NRs?
\end{enumerate}

%
 %

To address the questions above, it is convenient to start by comparing X and XX wavefunctions in NCs, NPLs and NRs.
In a rigid 3D crystal, the envelope function of a Wannier exciton has a hydrogenic form, $\psi_X=e^{-r_{eh}/a_B^{3D}}$.
Here, $r_{eh}$ is the electron-hole distance and $a_B^{3D}$ the Bohr radius. 
In atomic units, the latter is set by $a_B^{3D}=\varepsilon_r/\mu$, with $\mu$ the reduced mass of X and $\varepsilon_r$ the relative dielectric constant.
The Bohr radius is a very useful magnitude to describe excitonic properties.
For example, the X binding energy is $\Delta_X^{3D}=1/(2 \mu (a_B^{3D})^2)$.
With decreasing dimensionality, the Bohr radius is known to decrease:
 $a_B^{2D}=a_B^{3D}/2$ and $a_B^{1D}\rightarrow 0$.\cite{MossAJP,AtawnahAIP} 
Consequently, binding energies are expected to increase: $\Delta_X^{2D}=4\Delta_X^{3D}$ and $\Delta_X^{1D}\rightarrow \infty$.
$a_B$ is also relevant to describe the influence of superradiance on radiative recombination rates.
In cuboids, dipolar transitions scale proportional to $(L/a_B)^N$, where $N$ is the number of weakly confined directions
and $L$ their length.\cite{PlanellesAP}
$a_B$ is also useful to determine the confinement regime of nanocrystals. 
A widespread rule of thumb is that a direction is strongly confined when its length is under the Bohr radius, $L<a_B$, and weakly confined otherwise.
This has implications on the physical factors that prevail in determining the X response, either kinetic energy or Coulomb correlations.

The problem is that $a_B$ is ill-defined in metal halide perovskite nanostructures.
The first reason is that confinement in NRs, NPLs and NCs is not in the strict 1D, 2D and 3D limit.
$\psi_X$ in these finite crystals departs from the hydrogenic expressions.\cite{PlanellesTCA}
Even if the wavefunction may preserve a $e^{-Z r_{eh}}$ term,\cite{MovillaPRB,ClimenteJPCL,TliliJPCc} 
the inverse of the coefficient $Z$ (a pseudo-Bohr radius) no longer corresponds to the most probable electron-hole distance.
The second reason is that $\varepsilon_r$ is no longer a local variable,
owing to polaronic and dielectric confinement effects.\cite{MovillaNSa,ClimenteJPCL}
For example, polaronic effects make $\varepsilon_r$ approach the static limit away 
from the polaron radius, and the high-frequency limit inside it.\cite{BaranowskiAEM,MovillaNSa}
Since $a_B^{3D}=\varepsilon_r/\mu$ (atomic units), $a_B^{3D}$ can be anywhere from $1.5$ to $5.4$ nm.

\begin{figure}[h]
\includegraphics[width=15.0cm]{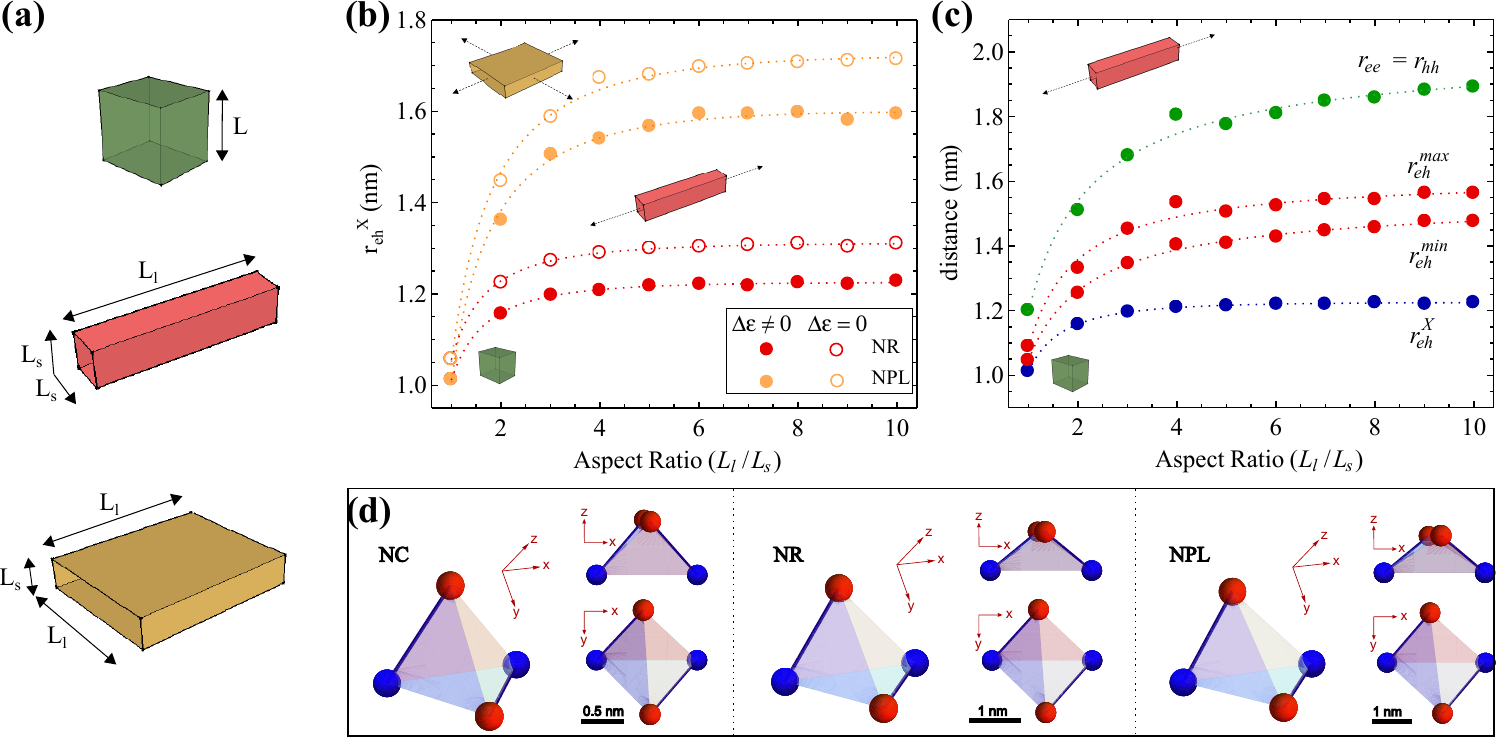}
\caption{
	Influence of shape on the spatial charge distribution of X and XX.
	(a) Schematic of the nanocrystal shapes under consideration.
	(b) Effect of anisotropy on the mean electron-hole distance of X in
	nanocrystals with $L_s=3$ nm..
	The weakly confined direction is influential unless the aspect ratio is large.
	The number of confined dimensions imposes significant differences in $r_{eh}^X$.
	The presence (solid dots, $\Delta \varepsilon \neq 0$) or absence (empty dots, $\Delta \varepsilon = 0$) 
	of dielectric mismatch plays a secondary role. 
	(c) Same but including XX carrier-carrier distances, in NRs.
	Electron-hole distances are enlarged as compared to X.
	(d) XX geometry as inferred from the distances.
	A distorted tetrahedron is formed regardless of the dimensionality.
	$x$, $y$ and $z$ are internal coordinates.
}
\label{fig_geo}
\end{figure}

To provide an estimate of the actual size of X in CsPbBr$_3$ nanostructures, 
we obtain the X and XX ground state wavefunctions using a variational effective mass
Quantum Monte Carlo method (see Computational Methods).
This is done for cubic NCs (edges of length $L$), NRs (short edges $L_s$, long edge $L_l$) 
and NPLs (short edge $L_s$, long edges $L_l$). 
A schematic of the systems under consideration is plotted in Figure \ref{fig_geo}(a).
From the random walks in the Monte Carlo integration, we extract the most likely carrier-carrier distances.
 Figure \ref{fig_geo}(b) shows the resulting electron-hole distances ($r_{eh}^X$) 
obtained for the X ground state in the presence 
(solid dots) and absence
 (empty dots) of dielectric mismatch.
%
The starting point is a cubic NC with $L=3$ nm.
This size approximately corresponds to the shortest edges reported 
in CsPbBr$_3$ NRs.\cite{ZhangJACS,ChenAFM,WenACS,ValeJPCL}
The calculated distance is then $r_{eh}^X \approx 1$ nm, somewhat smaller than (half) the NC size.
We then fix $L_s=3$ nm and stretch one or two long edges ($L_l$) to form a quasi-1D NR (red dots) 
or a quasi-2D NPL (orange dots).
Weakening the confinement increases $r_{eh}^X$, but does so differently depending on the dimensionality.
Let us consider the case with dielectric confinement (solid dots).
For NPLs (orange solid dots), the electron-hole separation saturates at $r_{eh}^{X,NPL}\approx 1.6$ nm,
while for NRs (red solid dots), it does so at $r_{eh}^{X,NR}\approx 1.2$ nm.
The smaller size in NRs as compared to NPLs is in line with expectations from the Bohr radius 
in infinite rigid crystals (see above), but the X shrinking is much smaller than in the strict 1D and 2D limits.

Because the X size in NRs is smaller than in NPLs, it stops feeling the long edge sooner. 
Thus, $r_{eh}^{X,NR}$ saturates at an aspect ratio $L_l/L_s \approx 4$ while $r_{eh}^{X,NPL}$ does so at $L_l/L_s \approx 6$.
Nonetheless, the long edge remains influential for aspect ratios well above $r_{eh}^X$.
For example, $r_{eh}^{X,NR}$ still feels $L_l$ for an aspect ratio of 2-3 ($L_l=6-9$ nm).
This is indicative that the $X$ wavefunction is actually spreading well beyond the most likely electron-hole distance.

All the observations described above hold in the absence of dielectric confinement (empty dots), 
even though in that case the distances are moderately longer because Coulomb attractions are weaker.

Electron-hole distances are substantially modified upon formation of a XX.
Figure \ref{fig_geo}(c) considers a NC evolving into a NR,
and compares the electron-hole distance inside a single X ($r_{eh}^X$, blue dots),
with that inside a XX ($r_{eh}^{min}$ and $ r_{eh}^{max}$, red dots, roughly corresponding to intra-
and inter-X distances), as well as electron-electron or hole-hole distances ($r_{ee}$, $r_{hh}$, green dots).
One can see that for all aspect ratios, $r_{eh}^{min} > r_{eh}^X$. 
That is, electron-hole distances increase upon the formation of XX.
 In other words, the interactions across the two X forming XX relax attractions within X.

From the carrier-carrier distances in Fig.~\ref{fig_geo}(c), 
we can infer the most likely XX molecular geometry.\cite{ClimenteJPCL} 
Figure \ref{fig_geo}(d) compares the spatial distribution of XX charges in a NC ($L=3$ nm), 
with that of a NPL and a NR with the same short edge but large aspect ratio ($L_l/L_s=10)$.
The result clearly shows that, despite difference in the length scale, 
the distribution is very similar in all cases, namely a distorted tetrahedron.
The same geometry was previously reported in CdSe and MAPbI$_3$ NPLs.\cite{MaciasNS,ClimenteJPCL}
It is interesting that the same distribution of charges is obtained in NCs, NPLs and NRs despite strong anisotropies.
If Coulomb interaction were the leading force in the system, in the NR one would expect a linear chain of charges
with alternating sign, while in the NPL one would expect a square with charges of the same sign sitting at opposite vertexes.\cite{SinghPRB}
If kinetic energy were the driving factor instead, the random motion of particles would make all carrier-carrier distances equal
and a perfect tetrahedral distribution would result.
The distorted tetrahedron points out at a competition between the two factors in all nanocrystals, 
with a dominant contribution of kinetic terms perturbed by Coulomb ones.\\

We next study the effect of dimensionality on the X radiative recombination rates.
Exciton radiative rates in the low-temperature limit are evaluated within the dipole approximation:\cite{PlanellesJPCc}
\begin{equation}
k_{X}^{rad} \propto \sum_{\alpha=x,y,z} f_{LF,\alpha}^2 | \langle 0 | p_\alpha | \Psi_X \rangle |^2.
\label{eq:k}
\end{equation}
Here, $f_{LF,\alpha}$ is the local field factor, accounting for the dielectric screening of the electromagnetic field 
polarized along the $\alpha$ direction, and $| \langle 0 | p_\alpha | \Psi_X \rangle |^2$ the interband transition element, 
describing the coupling between the X ground state and the closed shell configuration.
Fine structure details are disregarded because bright-dark state coupling is weak in CsPbBr$_3$ nanostructures.\cite{TamaratNC}
Both terms in $k_X^{rad}$ are sensitive to anisotropy, and provide independent contributions which differ for NC, NPLs and NRs.
$f_{LF,\alpha}$ has been shown to favor directional absorption and emission in NRs\cite{KamalPRB} and NPLs.\cite{ScottNN}
Analytical expressions for this effect can be obtained by approximating the nanocrystal geometry as ellipsoids.
This is a reasonable approximation because confinement makes the exciton charge density avoid the corners of cuboids,
which justified its use in previous studies of NPLs.\cite{ScottNN}
The local field factor is then given by\cite{SihvolaJNm}
%
	$f_{LF,\alpha}=\varepsilon_{out} / \left( \varepsilon_{out}+N_{\alpha}(\varepsilon_{\infty}-\varepsilon_{out}) \right)$ ,
%
\noindent with $N_{\alpha}$ the corresponding depolarization factor, $\varepsilon_{\infty}$ the CsPbBr$_3$ 
high-frequency dielectric constant and $\varepsilon_{out}$ the dielectric constant of the surrounding medium.
$N_\alpha$ is calculated through simple integrals\cite{SihvolaJNm}. In isotropic structures, $N_x=N_y=N_z=1/3$.
In anisotropic structures, $N_\alpha$ decreases along the weakly confined directions and increases otherwise.
This implies that $f_{LF,\alpha}$ approaches one (no field screening) along the weakly confined direction,
which may translate into a net enhancement of the $k_{X}^{rad}$ through Eq.~(\ref{eq:k}).

As for the dipole matrix element, factorizing envelope and Bloch parts as 
$\Psi_X=\psi_X \, |u_{cb}\rangle \, |u_{vb}\rangle$ 
leads to $\langle 0 | p_\alpha | \Psi_X \rangle = \langle 0 | \delta(\mathbf{r}_e, \mathbf{r}_h) | \psi_X \rangle \, K_\alpha$,
with $K_\alpha=\langle u_{cb}|p_\alpha|u_{vb}\rangle $ a Kane-like factor. 
Since the Bloch functions of conduction and valence bands are nearly isotropic,\cite{SercelNL}
we approximate $K_x=K_y=K_z$. The influence of nanocrystal dimensionality on $\langle 0 | p_\alpha | \Psi_X \rangle$
is then set by the envelope part only, $S_{eh}=\langle 0 | \delta(\mathbf{r}_e, \mathbf{r}_h) | \psi_X \rangle$,
i.e. the electron-hole overlap integral. We have shown in Fig.~\ref{fig_geo} that different nanocrystal shapes 
lead to different $r_{eh}^X$ values, which again should impact $k_X^{rad}$.

\begin{figure}[h]
\includegraphics[width=15.0cm]{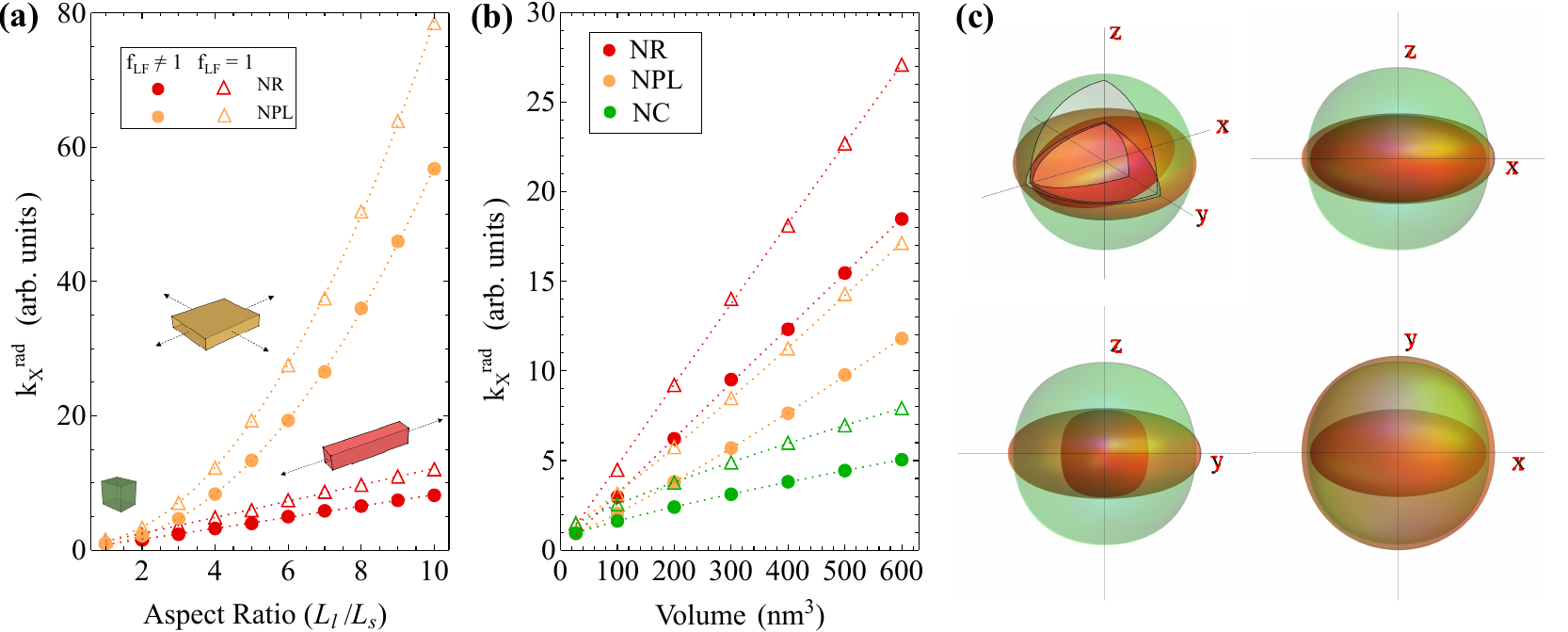}
\caption{
	Influence of shape on the X recombination rate.
	(a) Effect of anisotropy, starting from a NC with $L=3$ nm.
	The aspect ratio has a greater influence on NPLs because of the larger volume.
	The local field factor is a relevant, yet secondary contribution.
	(b) Radiative rate for nanocrystals of different shape but equal volume.
	NRs give the fastest recombination, because of the smaller electron-hole distance. 
	(c) Surfaces containing 90\% of excitonic electron probability 
	when the hole is at the center of the nanocrystal.
	Red, orange and green represent NR, NPL and NC.
	In all cases, $L_s=3$ nm and $V=500$ nm$^3$.
	The smaller the dimensionality, the smaller the volume of charge distribution,
	which explains the relative radiative rates.
}
\label{fig_rad}
\end{figure}

To elucidate the relative contributions of $f_{LF}$ and  $S_{eh}$ in different types of nanocrystals,
in Figure \ref{fig_rad}(a) we calculate $k_X^{rad}$ for a NC with $L=3$ nm, which evolves into a NR or a NPL.
Dots (triangles) correspond to simulations including (excluding) the influence of $f_{LF}$.
In both sets of simulations, $k_X^{rad, NPL}$ and $k_X^{rad, NR}$ increase as the long axis is elongated
(aspect ratio increases), with $k_X^{rad,NPL} > k_X^{rad,NR}$.
This is a consequence of superradiance\cite{ZhuNAT}, which for cuboids scales as $(L/a_B)^N$, 
where $N$ is the number of weakly confined directions.\cite{PlanellesAP}
%
The anisotropic dielectric screening of the electromagnetic field does not change this trend,
as the local field factor quenches NR and NPL recombination alike.
The effect, even if relevant, is secondary as compared to superradiance.

The faster emission of NPLs as compared to NRs observed in Fig.~\ref{fig_rad}(a) is at odds with
current experimental data, which points at NRs as the fastest emitting CsPbBr$_3$ nanocrystals.\cite{ZhuNL}
 In our simulations, the exciton delocalization volume corresponds to the entire nanocrystal,
which is an ideal limit. Thus, the disagreement is likely indicating that X in NPLs are in practice localized within a subregion of the
available space, as is the case in CdSe NPLs.\cite{GeiregatLSA} 
For an alternative comparison, we next consider the volume of X delocalization is the same for
all three structures. Figure \ref{fig_rad}(b) shows $k_X^{rad}$ calculated in NCs, NPLs and NRs as 
a function of their volume. Under these conditions, $k_X^{rad,NR} > k_X^{rad,NPL} > k_X^{rad,NC}$,
the trend being robust against inclusion or exclusion of local field factor effects.

The hierarchy of relaxation rates in Fig.~\ref{fig_rad}(b) is set by the electron-hole distances. 
For a cuboid with volume $V$, $S_{eh}^2 \propto V/a_B^3$.\cite{PlanellesAP}
Qualitatively, we can identify $a_B$ with $r_{eh}^X$. 
 Because the lower the dimensionality, the smaller $r_{eh}^X$ (recall Fig.~\ref{fig_geo}(b)),
 $S_{eh}$ is magnified. 
 To better illustrate this point, we go beyond the scalar value $r_{eh}^X$ and 
 calculate isosurfaces of probability from $\psi_X$, containing 90\% of the electron charge density
 when a hole is fixed at the center of the nanocrystal.
 Figure \ref{fig_rad}(c) compares the isosurfaces in a NC (green), a NPL (orange) and a NR (red) with volume $V=500$ nm$^3$.
 One can see that the electron is more concentrated around the hole position in the NR.
 This is a consequence of the strong spatial confinement in the transverse directions, along with
 the enhanced Coulomb interaction along the longitudinal one.
 Both effects force electron and hole to be closer together. 


\begin{figure}[h]
\includegraphics[width=7.0cm]{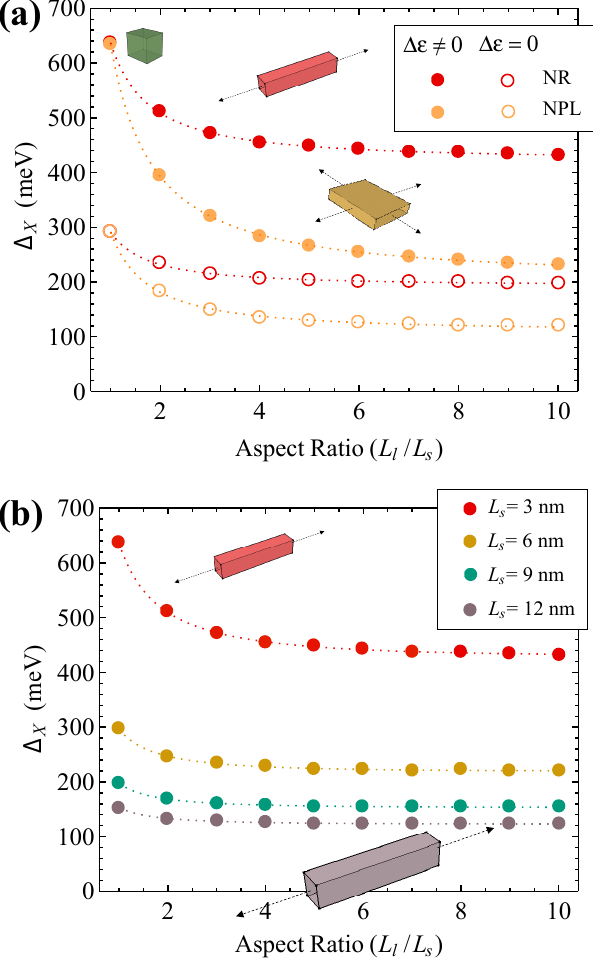}
\caption{
	Influence of shape and size on the exciton binding energy, $\Delta_X$.
	In general, the strong confinement directions play a leading role.
	(a) Effect of anisotropy on different shapes (NCs, NPLs and NRs)
	whose shortest edge is $L_s=3$ nm. 
	Solid (empty) dots: dielectric mismatch is (not) taken into account.
	The weakly confined direction is influential unless aspect ratio is large.
	The number of confined directions imposes substantial differences in $\Delta_X$.
	(b) Effect of anisotropy on NRs with different size of the cross-section.
	The impact of longitudinal confinement becomes weaker with increasing cross-section.
}
\label{fig_EbX}
\end{figure}

The exciton size is also key to understand the influence of dimensionality on the X binding energy.
To address this point, we follow the standard definition, $\Delta_X = (E_e + E_h) – E_X$, 
where $(E_e+E_h)$ is the energy of the non-interacting electron-hole pair and $E_X$ that of the interacting exciton.
Figure \ref{fig_EbX}(a) shows $\Delta_X$ calculated for nanocrystals with different shape,
from a NC with $L=3$ nm to NRs and NPLs. 
%
%
%
The qualitative trends are clearly connected with those of $r_{eh}^X$ (Fig.~\ref{fig_geo}(b)).
Thus, $\Delta_X^{NPL} < \Delta_X^{NR}$, as expected from $r_{eh}^{X,NPL}>r_{eh}^{X,NR}$.
Also, $\Delta_X^{NR}$ reaches the asymptotic limit sooner than $\Delta_X^{NPL}$.
The magnitude of the changes in $\Delta_X$ with the dimensionality is worth noting. 
For the size under consideration ($L_s=3$ nm), each confined dimension adds $\sim 33$\% to $\Delta_X$.
 That is, $\Delta_X^{NR} \approx 2/3\, \Delta_X^{NC}$ and $\Delta_X^{NPL} \approx 1/3\, \Delta_X^{NC}$.
 The same proportions hold in the absence or presence of dielectric confinement 
 (cf. solid and empty dots). 
 %
%

 The analysis of shape is complemented with an analysis of size.
In Figure \ref{fig_EbX}(b) we represent $\Delta_X$ for NCs of different initial edge lengths,
evolving into NRs. Again, one observes a saturation of $\Delta_X$ with the aspect ratio, 
but the larger the NR cross-section ($L_y \cdot L_z$), the sooner it takes place.
Also, the NR saturation values are gradually closer to that of the initial NC, $\Delta_X^{NR} \sim \Delta_X^{NC}$,
as expected in the bulk limit.

It follows from Figure \ref{fig_EbX} that $\Delta_X$ is mainly set by the strongly confined directions (if any). 
 Weakly confined directions can however provide a valuable tuning knob for small aspect ratios,
and dimensionality plays an important role. 
 A recent experimental work for NCs and NRs also suggested that the strong confinement direction
 is critical in determining $\Delta_X$.\cite{OrielNL}
 In fact, by fitting the static absorption spectra to an Elliott model, the authors inferred
 that $\Delta_X$ has an inverse cube dependence on $L_s$ only, 
 no matter the geometry. A unified expression, $\Delta_X= \Delta_X^{bulk} + k/L_s^3$,
 with $\Delta_X^{bulk}$ and $k$ fitting parameters, applied to both NCs and NRs. 
 Our results show that this is not the general case. In strongly confined nanostructures,
 e.g. those in Fig.~\ref{fig_EbX}(a), $\Delta_X$ depends not only on $L_s$ but also on the
 dimensionality. The fact that Ref.~\cite{OrielNL} gives a shared expression for NCs and NRs
 may be a consequence that the NRs they measure have moderate-to-weak confinement ($L_s \geq 6$ nm),
 where the effect of $1/L_s^3$ falls under the noise of a multi-parameter fitting procedure (Elliott model).

Another relevant conclusion from Figure \ref{fig_EbX}(a) is that dielectric confinement has a major influence on 
the excitonic structure of CsPbBr$_3$ nanostructures (cf. solid and empty dots).  
In the presence of short edges ($L_s=3$ nm), it can double $\Delta_X$ of NPLs, NRs and NCs alike.  
In absolute terms, because $\Delta_X^{NC}>\Delta_X^{NR}>\Delta_X^{NPL}$, doubling the binding energy 
implies a stronger effect of dielectric confinement with the number of confined dimensions.\\
%

\begin{figure}[h]
\includegraphics[width=8.0cm]{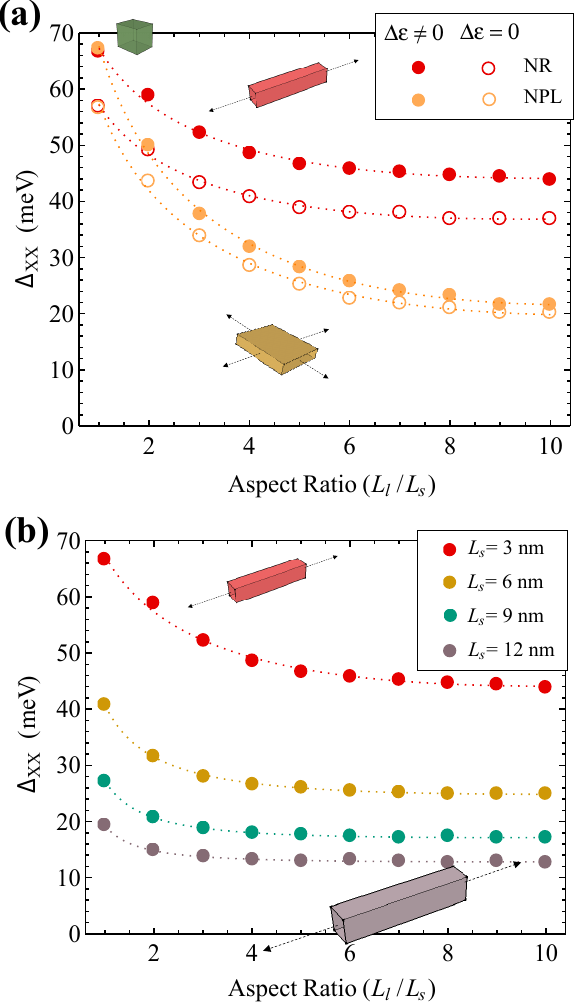}
\caption{
	Same as Fig.~\ref{fig_EbX}, but for biexciton binding energy, $\Delta_{XX}$.
	The effect of shape and size is the same as for $\Delta_X$, 
	despite the latter being mostly a charge-charge interaction 
	and the former a dipole-dipole one.
}
\label{fig_EbXX}
\end{figure}

The XX binding energy is another magnitude of practical interest, 
e.g. to achieve optical gain in lasing devices, where decoupling
of the emission frequencies of X and XX is desirable.
The XX binding energy is defined as $\Delta_{XX} = 2E_X - E_{XX}$,
with $E_X$ the exciton ground state energy and $E_{XX}$ the biexciton one.
Figure \ref{fig_EbXX} shows that the impact of shape, size and dimensionality 
on $\Delta_{XX}$ is similar to that on $\Delta_X$.
For nanocrystals with short edges of $L_s=3$ nm (Fig.~\ref{fig_EbXX}(a)),
the NR saturates at $\Delta_{XX}^{NR}\approx 2/3 \Delta_{XX}^{NC}$,
while the NPL does so at $\Delta_{XX}^{NPL}\approx 1/3 \Delta_{XX}^{NC}$.
The saturation takes place sooner for NRs than for NPLs, which
again suggests a smaller XX radius.
Fig.~\ref{fig_EbXX}(b) shows that the strong confinement directions are also
chiefly responsible for $\Delta_{XX}$ in NRs. 
Qualitatively, the main difference between $\Delta_{XX}$ and $\Delta_X$ 
is the influence of dielectric confinement. By comparing Fig.~\ref{fig_EbXX}(a)
and \ref{fig_EbX}(a), one can see the effect on $\Delta_{XX}$ is much smaller.
This is because $\Delta_X$ includes direct Coulomb plus correlation terms,
while $\Delta_{XX}$ includes correlation terms only, as direct Coulomb
terms of $E_{XX}$ and $E_X$ cancel out.
The similar influence of dimensionality on $\Delta_{X}$ and $\Delta_{XX}$
is an interesting result, because the different nature of electron-hole
and X-X interactions did not grant this result. 

A comment is however worth on the absolute values of 
$\Delta_X$ and $\Delta_{XX}$ we report.  
$\Delta_X$ is highly sensitive to the polaron radius. 
The polaron radii we consider ($l_e=l_h=3$ nm) are taken from bulk,\cite{BaranowskiACSener}
but extreme quantum confinement may alter the lattice structure and phonon modes, 
as observed e.g. in two-monolayer NPLs.\cite{MahmoudJPCL}
 Inferring the polaron radius in confined systems from theoretical considerations
alone is beyond reach of effective mass theory. In bulk, $l=\sqrt{\hbar^2/2\,m^b\,E_{LO}}$,
where $m^b$ is the bare mass of the carrier, and $E_{LO}$ the longitudinal optical phonon energy.\cite{BaranowskiACSener}
Because strong quantum confinement induces band non-parabolicity, and hence increase $m^b$\cite{BenchamekhPRB},
one may expect $l$ to decrease. 
A direct comparison of our model with experimentally measured $\Delta_X$ data 
could help verify this trend and determine effective polaron radii in such a regime, 
but this is handicapped by the difficulty of obtaining $\Delta_X$ experimentally:
measurements are indirect, and different methods give rise to very different 
$\Delta_X$ values.\cite{RodaNL} 
 Trion binding energies offer more accurate experimental data,
as they can be extracted directly from spectral shifts in the photoluminescence, 
but we are not aware of such data being available for CsPbBr$_3$ NPLs or NRs.
For this reason, our analysis is restricted to structures with $L \geq 3$ nm.
On the other hand, there is gathering evidence that, in CsPbBr$_3$ nanocrystals,
XX feel weaker dielectric screening than X.\cite{ZhuAOM,MovillaPRB} 
This unusual behavior has been ascribed to the slower exciton dynamics in XX.\cite{PlanellesPRB}
Anticipating the dielectric constant felt by XX is beyond reach of effective mass models,
so we take the same dielectric constants for X and XX, which leads to underestimated $\Delta_{XX}$
as compared to experiments.\cite{OrielNL}
Yet, the relative impact of shape and size in Fig.~\ref{fig_EbXX} remains valid for
an analysis of trends.

\begin{figure}[h]
\includegraphics[width=15.0cm]{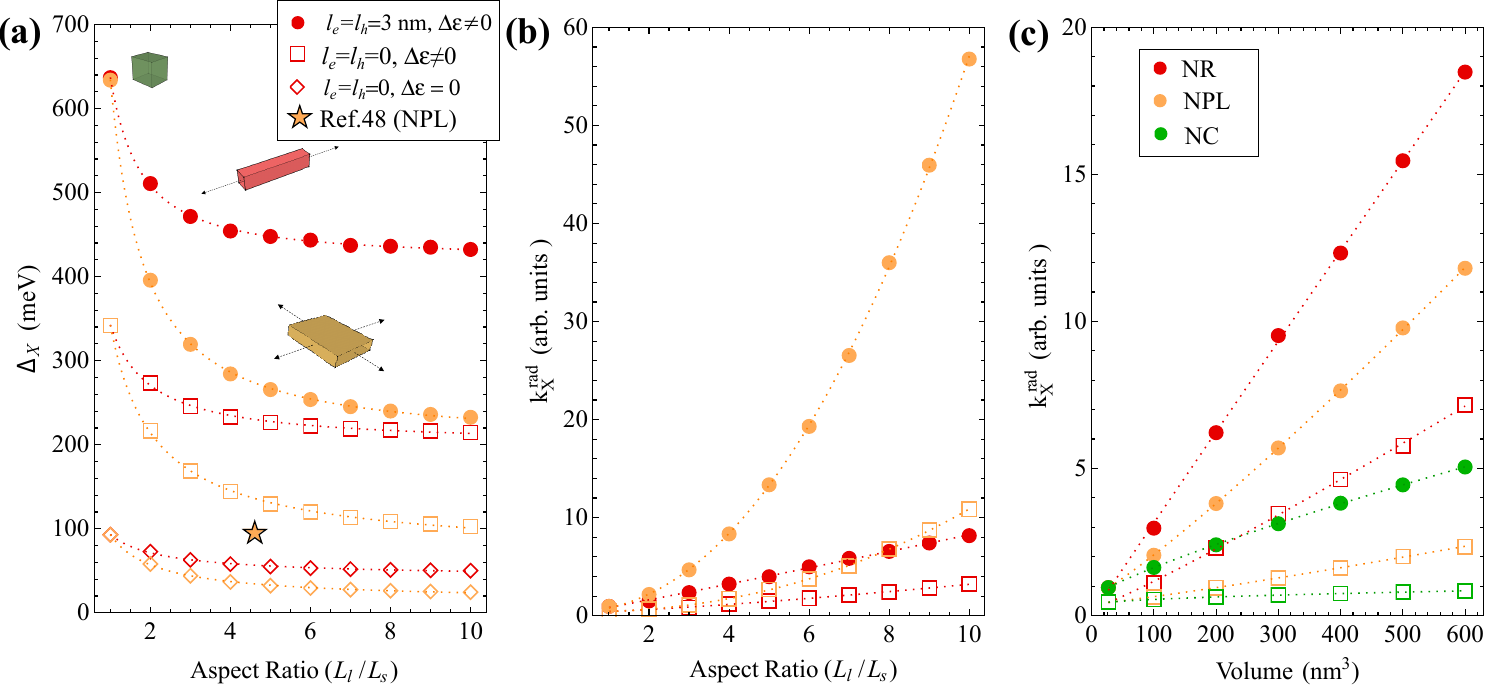}
\caption{
	Influence of polaronic radius and dielectric mismatch on excitonic properties.
(a) Exciton binding energies. Dots: full calculation (as in Fig.~\ref{fig_EbX}).
Squares: zero polaron radius. Rhombs: zero polaron radius and no dielectric confinement. 
Star: experimental point taken from Ref.~\cite{BohnNL} for a nanoplatelet with Ls=3 nm.  
The calculated binding energies approach the experimental magnitude when the polaron radius is reduced as compared to the bulk limit, and/or dielectric confinement is weakened. In all cases, however, nanorods show systematically greater binding energies than nanoplatelets.
(b,c) Exciton radiative rates of nanorods (red symbols), nanoplatelets (orange symbols) and nanocubes (green symbols).  Switching off polaronic effects (squares) reduces the radiative rates but preserves the hierarchy observed in the full calculation. 
}
\label{fig_refs}
\end{figure}

To test the robustness of the qualitative conclusions obtained so far, 
in Figure \ref{fig_refs} we plot $\Delta_X$ and $k_X^{rad}$ in the 
full calculations (dots), as compared to those in the absence of polaronic effects
(squares) and dielectric confinement effects (rhombs).
As a reference point, in Fig.~\ref{fig_refs}(a) the experimentally measured
value of $\Delta_X$ for NPLs with $L_s=3$ nm is given.\cite{BohnNL}
Clearly, the full calculations largely overestimate this energy.
Reducing the polaron radius (squares) improves agreement, 
and so does reducing the dielectric mismatch (rhombs).
It is uncertain to which extent $l_e,\,l_h$ and $\Delta \varepsilon$ are reduced
in the experiments, but the relevant observation is that in all three
sets of simulations, the hierarchy of $\Delta_X$ is set by the dimensionality.
 Likewise, Figs.~\ref{fig_refs}(b,c) show that suppressing polaronic effects reduces
$k_X^{rad}$ but preserves the qualitative trends established by dimensionality.

 From the results presented above, we can propose answers to the questions we had placed
 in the beginning of this study: 

\begin{enumerate}[label=(\roman*)]
	\item The strong confinement direction is the main factor determining $\Delta_X$ and $\Delta_{XX}$ in
		CsPbBr$_3$ nanocrystals, as pointed out in Ref.~\cite{OrielNL}.
		However, we hypothesize that the simple relations of Ref.~\cite{OrielNL}, 
		hold only within experimental uncertainty.
		In general, $\Delta_X$ and $\Delta_{XX}$ are dependent of the 
		dimensionality and on the length of the weakly confined directions.
		These provide a significant modulation unless $L \gg r_{eh}^X$. 
	\item Dielectric confinement provides a major contribution to $\Delta_X$ (up to 50\% in nanocrystals with $L=3$ nm),
		and much less to $\Delta_{XX}$. This is because the former involves direct Coulomb interactions, which
		in the latter largely cancel out.
	\item Strongly confined directions determine the strength of correlations in anisotropic nanocrystals.
		These are reflected in the electron-hole distances,
		$r_{eh}^X$.
		The distances decrease (and hence correlations increase) from 3D to 2D and to 1D.
	\item Exciton recombination in NRs is faster than that in NPLs or NCs of the same volume,
		because strong confined directions lead to shorter $r_{eh}^X$ distances.
		The anisotropic screening of the electromagnetic field is a secondary (yet sizable) contribution.
		The recombination in NPLs could be faster for large enough areas, but X localization is likely taking place in these structures.
		These results explain the observations in Ref.~\cite{ZhuNL}.
	\item The spatial distribution of charges in XX is a distorted tetrahedron, regardless of the shape of the nanocrystal (NC, NPL or NR).
		This geometry reflects the competition between kinetic and Coulomb terms, with the former prevailing in all cases.
\end{enumerate}

\section{Computational methods}

We calculate the X and XX ground-state energies ($E_X$ and $E_{XX}$) and wavefunctions ($\psi_X$ and $\psi_{XX}$)
in cuboidal CsPbBr$_3$ NCs using single-band $k \cdot p$ Hamiltonian for conduction and valence bands. 
Quantum confinement is accounted for by means of a 3D potential. 
Dielectric confinement, by means of the image charge method.\cite{MovillaNSa}
X and XX correlations, by means of variational Quantum Monte Carlo method.\cite{PlanellesCPC,MaciasNS}
Electron-hole interaction with the polar lattice (polaron formation), 
by means of a renormalized Haken potential.\cite{BaranowskiAEM}
A detailed description of the resulting model can be found in Ref.~\cite{MovillaPRB}. 

 Two relevant aspects are worth stressing here. 
 The first one is related to the trial wavefunction. 
 Our trial wavefunctions are products of non-interacting particle-in-a-box states and correlation factors.\cite{MovillaPRB} 
 In the present study, electron-hole attraction is captured by a Slater correlation factor 
 $e^{-Z_X s_{eh}}$, with $s_{eh}=\{(x_e-x_h)^2 + b [(y_e-y_h)^2+(z_e-z_h)^2]\}^{1/2}$, 
 and $Z_X$ and $b$ the parameters to optimize variationally. 
$Z_X$ denotes the strength of the electron-hole correlation, whereas $b$ captures its anisotropy. 
This makes the trial function flexible enough to describe excitons in cuboids with very different aspect ratios. 
Thus, the optimized value $b_{opt}<1$ for NRs,  $b_{opt}>1$ for NPLs, and $b_{opt}=1$ for NCs. 

The second aspect relates to dielectric confinement.
To account for the dielectric mismatch with the surrounding medium, interactions are calculated using a modified image-charge 
expansion\cite{MovillaNSa} based on Takagahara's expressions for rigid semiconductor cuboidal structures.\cite{TakagaharaPRB}. 
 Bolcatto and Proetto found that the Takagahara expansions do not behave properly in certain limits.\cite{BolcattoPRB}
 A plausible reason is that the boundary conditions that safely apply to infinite dielectric surfaces may not be freely extrapolated to finite or orthogonal surfaces without generating pathological singularities or discontinuities. 
However, the authors compared the results of the Takagahara potentials with those obtained through the calculation of the 
polarization charges induced at the dielectric interface, and found that the difference between both approaches was nearly 
independent of the source and test charges within the cuboidal dots. 
Following a similar strategy, we have verified that, within the cuboids, our modified image-charge expansion\cite{MovillaNSa} 
matches (up to a nearly constant value) the potential calculated by solving Poisson's equation with proper Dirichlet boundary conditions. 
Since the first method is computationally much more efficient, we use it after calculating through the second the 
corresponding correction for each geometry under study.\\

From the optimized wavefunctions, the XX geometry is analyzed by calculating the most probable location of electrons and holes 
from the internal coordinates recorded during the Monte Carlo random walks, as these rely on the Metropolis algorithm. 
To capture the potential asymmetry of the interactions, which may yield inter-exciton distances larger than intra-exciton ones, 
we calculate the average distances between a hole and the closer ($\langle r_{eh}^{min} \rangle$) or farther ($\langle r_{eh}^{max} \rangle$) electron. 
Finally, to eliminate the inherent statistical bias of this definition, results are renormalized against an uncorrelated system ($Z=0$),
see Ref.~\cite{ClimenteJPCL} for further details.\\







CsPbBr$_3$ NCs are described using the material parameters extracted in Ref.~\cite{BaranowskiACSener}. 
Thus, polaronic electron and hole masses are
 $m_e=m_h=0.316\,m_0$, with $m_0$ the free electron mass,
 while polaron radii are $l_e = l_h = 3$ nm.
Relative static and dynamic dielectric constants inside the NC 
are $\varepsilon_{s} = 16$ and $\varepsilon_{\infty} = 4.5$.
  The dielectric constant of the surrounding matrix is set to 
 $\varepsilon_{out} = 2.56$,
 as corresponding to polystyrene,\cite{ZhuAOM}
 when dielectric mismatch is taken into account ($\Delta \varepsilon \neq 0$).
 When it is disregarded ($\Delta \varepsilon = 0$), no image charges are computed and $\varepsilon_{out}$ is irrelevant. 
It has been shown that, when inserted in our model, these parameters 
give good estimates of X and trion energies in CsPbBr$_3$ quantum dots.\cite{MovillaPRB}


\acknowledgement
We acknowledge support from Grant No. PID2024-162489NB-I00, funded by Ministerio de Ciencia, Innovación y Universidad
(MICIU /AEI /10.13039/501100011033 / FEDER, EU) and Universitat Jaume I through project GACUJIMA-2025-14.

\bibliography{bib_pv}

\end{document}